\documentclass[useAMS,usenatbib]{mn2e}

\usepackage{graphicx}
\newcommand{\chandra}{{\it Chandra}}

\newcommand{\xmm}{{\it XMM-Newton}}

\title[0.75\,mHz oscillation in V4743\,Sgr is not multiple]{Counter evidence against multiple frequency nature of 0.75\,mHz oscillation in V4743\,Sgr}

\author[A. Dobrotka and J.-U. Ness ]{
A. Dobrotka$^1$\thanks{E-mail: andrej.dobrotka@stuba.sk},
J.-U. Ness$^2$\\
$^1$Advanced Technologies Research Institute, Faculty of Materials Science and Technology in Trnava, Slovak University of Technology\\
in Bratislava, Bottova 25, 917 24 Trnava, Slovakia\\
$^2$XMM-Newton Science Operations Center, European Space Astronomy Center, Camino Bajo del Castillo s/n, Urb. Villafranca del\\
Castillo, 28692 Villanueva de la Ca\~nada, Madrid, Spain\\
}

\begin{document}

\date{Accepted ???. Received ???; in original form \today}

\pagerange{\pageref{firstpage}--\pageref{lastpage}} \pubyear{2017}

\maketitle

\label{firstpage}

\begin{abstract}
All X-ray light curves of nova V4743\,Sgr (2002), taken during and after outburst, contain a 0.75\,mHz periodic signal that can most plausibly be interpreted as being excited by the rotation of the white dwarf in an intermediate polar system. This interpretation faces the challenge of an apparent multi-frequency nature of this signal in the light curves taken days 180 and 196 after outburst. We show that the multi-sine fit method, based on a superposition of two sine functions, yields two inherently indistinguishable solutions, i.e. the presence of two close frequencies, or a single signal with constant frequency but variable modulation amplitude. Using a power spectrum time map, we show that on day 180, a reduction of the modulation amplitude of the signal coincides with a substantial overall flux decline while on day 196, the signal was only present during the first half of the observation. Supported by simulations, we show that such variations in amplitude can lead to false beating that manifests itself as a multiple signal if computing a periodogram over the full light curve. Therefore, the previously proposed double frequency nature of both light curves was probably an artefact while we consider a single signal with frequency equal to the white dwarf rotation as more plausible.
\end{abstract}


\section{Introduction}
\label{introduction}

Cataclysmic variables are interacting binary systems, consisting of a white dwarf primary and a late main sequence star (see e.g. \citealt{warner1995} for a review).  Accretion takes place from the cool companion to the white dwarf causing accumulation of hydrogen-rich material onto the white dwarf surface. If the accretion rate is in the range between $(1-4)\times 10^{-7}$\,M$_\odot$\,yr$^{-1}$, steady nuclear burning can be established, thus hydrogen is fused to helium at the accretion rate upon arrival on the white dwarf surface (\citealt{vandenheuvel1992}). If the accretion rate is lower, then slow accumulation of hydrogen-rich material leads to building up a hydrogen-rich surface layer that ignites explosively in a thermonuclear runaway when ignition conditions are reached. A critical amount of mass of $10^{-6}-10^{-4}$\,M$_\odot$ (depending on white dwarf mass) is needed for such an explosion which are commonly known as Classical Nova outburst. Subsequently, within hours, the white dwarf is engulfed in an envelope of optically thick material that is driven away from the white dwarf by radiation pressure. The nova is first bright in optical, but the peak of the spectral energy distribution shifts to higher energies as the mass ejection rate (and thus the opacity) decreases. As a consequence, the expanding envelope is becoming less opaque, and successively hotter layers become visible (see, e.g., \citealt{bode2008}). A few weeks to months after outburst, the nova becomes bright in X-rays with a dominant supersoft blackbody-like spectral component.

The nova V4743\,Sgr was discovered in September 2002 by \cite{haseda2002}. \citet{ness2003} found large-amplitude variability in \chandra\ data taken 180 day after the maximum brightness with a period of $\sim 22$ minutes (0.75\,mHz) with clear overtones of this signal in the periodogram. During this observation a strong decline in X-ray brightness was observed with a simultaneous spectral change from a continuum spectrum to emission lines.

More detailed period analyses of the \chandra\ and \xmm\ X-ray data taken 196 days after outburst were presented by \cite{leibowitz2006}. The large-amplitude variations albeit with lower amplitude were also found in the \xmm\ light curve, while the main feature at 0.75\,mHz showed two peaks leading to an interpretation of a multiple signal. The authors confirmed the double nature also in the earlier \chandra\ data taken on day 180. Moreover, at least 6 and 12 additional frequencies were identified on days 180 and 196, respectively. \cite{leibowitz2006} suggested that the main feature at 0.75\,mHz is related to the white dwarf rotation and that the other observed frequencies are produced by non-radial white dwarf pulsations.

\cite{kang2006} presented CCD unfiltered optical photometry analyses from observations taken one and three years after the outburst. They detected two periods of 6.7\,h and $\sim 24$ minutes. The authors attributed the longer signal to the orbital period of the underlying binary system. Assuming that the 22 minute signal present in X-ray (\citealt{ness2003}) is the rotation period of the central white dwarf, the longer optical period of 24\,min was interpreted as the beat period between the orbital and rotation period of the white dwarf, respectively. This suggests an intermediate polar nature of the V4743\,Sgr binary.

\citet{dobrotka2010} performed a complex timing study of all X-ray data taken up to 3.5 years after maximum brightness using two different period analysis methods. They confirmed that the \chandra\ data taken 180 days and \xmm\ data taken 196 days after outburst can be modulated with two signals with close frequencies as detected previously by \citet{leibowitz2006}. In later observations only a single signal was detected supporting the intermediate polar interpretation. Other unstable or multiple significant signals are compatible with pulsations.

\citet{zemko2016} reanalysed the latest X-ray observations taken 2 and 3.5 years after maximum brightness together with a new SALT optical spectrum taken 11.5 years after outburst. Besides the same procedure as performed by \citet{dobrotka2010}, the authors performed separate timing studies for a hard and a soft band. While they found the 0.75\,mHz signal in both bands at 2 years after outburst, they were only able to detect it in the hard band at 3.5 years after outburst. The X-ray spectra from both observations had the characteristics in common with known intermediate polars, with a hard thermal plasma fitted with a partially covering absorber. The earlier spectrum had also a supersoft blackbody-like component possibly originating from the polar regions irradiated by an accretion column. The authors emphasised the intermediate polar nature of the nova V4743\,Sgr.

Apparently the multifrequency nature of the oscillations in nova V4743\,Sgr is a challenge to the interpretation of a modulation arrising from the rotation of the white dwarf. We have revisited the two X-ray light curves with multiple signals, taken 180 and 196 days after maximum brightness, in order to elucidate whether the data are modulated with a single or multiple signals.

\section{Observations}

In this work we reanalyse some of the data studied in \citet{dobrotka2010}. We focus on the \chandra\ LETGS light curve (ObsID 3775) taken 180.39 days (hereafter day 180) after maximum and the \xmm\ EPIC-pn light curve (ObsID 0127720501) obtained 196.14 days (hereafter day 196) after the maximum brightness. All details about observations and data reduction are summarised in Table~1. We also refer to Section~2 in \citet{dobrotka2010}.

\section{Timing analysis}

\subsection{Method}
\label{method}

For timing analysis we use two different methods. One is to compute a standard 1-D periodogram (power vs. frequency) over the entire (detrended) light curves using the Lomb-Scargle algorithm by \cite{scargle1982} as in \citet{dobrotka2010}. This approach yields a representation of the overall light curve behaviour and is useful when dealing with light curves that are modulated with a discrete coherent and stable signal. As a visualisation of variations in signal strength and/or frequency during an observation, we display a series of periodograms as in \citet{ness2015} to form a time map. We divide a light curve into equally spaced, overlapping, subsets, and compute a 1-D (Lomb Scargle) periodogram for each of them. The subsets overlap with 50\%, i.e., for a duration of each subset of $\Delta\,t$, the periodograms are calculated for time intervals of 0 - $2\Delta\,t$, $\Delta\,t$ - $3\Delta\,t$, $2\Delta\,t$ - $4\Delta\,t$, etc. All 1-D periodograms are tagged together to form a 2-D time map.

\subsection{Analysis}

\begin{figure}
\resizebox{\hsize}{!}{\includegraphics[angle=0]{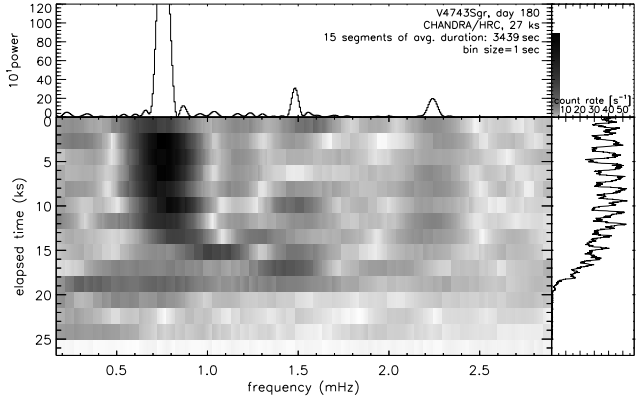}}
\caption{Time map of the \chandra\ observation from day 180. The whole light curve is divided into 15 equally spaced subsets in order to calculate individual periodograms. The light curve is added for direct comparison in the right panel. The top panel shows the overall periodogram and two selected periodograms calculated from subsets marked as red a blue shaded areas in the right panel.}
\label{time_map_day180}
\end{figure}

\begin{figure}
\resizebox{\hsize}{!}{\includegraphics[angle=0]{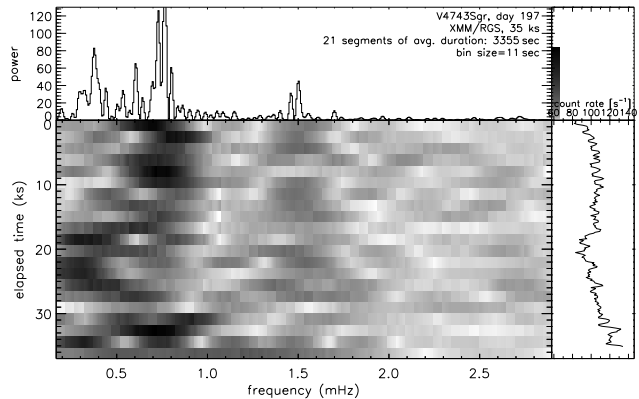}}
\caption{The same as Fig.~\ref{time_map_day180} but for day 196. The whole light curve is divided into 20 equally spaced subsets.}
\label{time_map_day196}
\end{figure}

The 1-D periodograms derived from the light curves taken on days 180 and 196 are presented in \citet{dobrotka2010} while the 2-D maps are in Figs.~\ref{time_map_day180} and \ref{time_map_day196} in this paper, respectively. Fig.~\ref{time_map_day180} shows a more or less constant main frequency until 13\,ks from observation start after which a strong flux decline started that may be accompanied by an increase of frequency until it becomes undetectable. In Fig.~ \ref{time_map_day196}, the main 0.75\,mHz signal seems slightly variable up to approximately 15-16\,ks with some weak signals possibly present up to 26\,ks.

Both time maps suggest that the 0.75\,mHz signal has changed both in terms of amplitude and frequency, making the analysis more complicated than previously assumed (\citealt{dobrotka2010}). The simultaneous change of frequency and X-ray flux on day 180 has to be taken into account in order to study whether the flux decrease may have mislead us to conclude a multifrequency signal. Moreover, the variations of the frequency and its later disappearance on day 196 can significantly influence the periodogram of the full light curve, and the multi-frequency nature of the main feature at 0.75\,mHz needs more detailed analysis.

For this re-analysis we used the same detrended light curve from day 180\footnote{Subtracting a fifth-order polynomial fit to the data.} and the non-detrended light curve from day 196 as in \citet{dobrotka2010}.

\subsection{Re-analysis of the light curve taken on day 180}
\label{fitting_day_180}

While for the observation on day 196, the multiple nature of the 0.75\,mHz signal was already detectable with standard approaches by \citet{leibowitz2006}, \citet{ness2003} only reported a single signal for day 180. We developed a more sensitive approach using a multi-sine fitting algorithm to investigate whether also on day 180 the signal actually consists of two components \citep{dobrotka2010}.

While the multi-sine fit was based on a superposition of two sine functions, the underlying basic mathematical relation
\begin{equation}
{\rm sin}(f_1) + {\rm sin}(f_2) = 2~{\rm cos}\left(\frac{f_1 - f_2}{2}\right)~{\rm sin}\left(\frac{f_1 + f_2}{2}\right)
\label{sine_equality_1}
\end{equation}
actually allows two solutions on the right-hand side. We performed direct fittings with the {\tt GNUPLOT}\footnote{http://www.gnuplot.info/} software using both solutions in the form of
\begin{equation}
a_1~{\rm sin}(f_1) + a_2~{\rm sin}(f_2) = a_0~{\rm cos}(F_1)~{\rm sin}(F_2),
\label{sine_equality_2}
\end{equation}
where
\begin{equation}
F_1 = \frac{f_1 - f_2}{2}~~{\rm and}~~F_1 = \frac{f_1 + f_2}{2}.
\label{sine_equality_3}
\end{equation}

\begin{figure}
\resizebox{\hsize}{!}{\includegraphics[angle=-90]{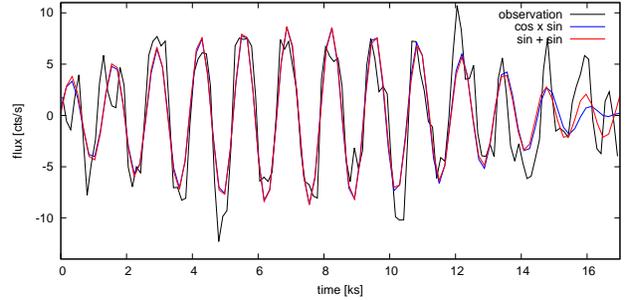}}
\caption{Detrended light curve from day 180. The data are averaged (every 250\,s) for clarity. The lines are different fits using equation~(\ref{sine_equality_2}).}
\label{day180_fits}
\end{figure}

The results are displayed in Fig.~\ref{day180_fits}, and fitted frequencies are summarised in Table~\ref{day180_fits_param}. Clearly, both fits yield very similar results with values (almost) satisfying equation~(\ref{sine_equality_3}). Therefore, two solutions are possible to explain the light curve from day 180, i.e. 1) two close signals as published by \citet{dobrotka2010}, and 2) a single frequency of 0.76\,mHz with a variable amplitude on a longer time scale of approximately 38.5\,ks (equivalent to a frequency of 0.026\,mHz). The latter is the cosine period. While only one antinode loop is visible on day 180, only half of the beating cycle (the light curve in Fig.~\ref{day180_fits}) should be compared to the cosine cycle, which is 17 vs. 19\,ks, respectively. These values match well as a rough estimate.
\begin{table}
\caption{Frequencies from fits to the day 180 data using eq.~(\ref{sine_equality_2}). The right column is recalculated from eq.~(\ref{sine_equality_3}) using frequencies from middle column.}
\begin{center}
\begin{tabular}{lcr}
\hline
\hline
frequency & from fits & from eq.~(\ref{sine_equality_3})\\
 & (mHz) & (mHz)\\
\hline
$f_1$ & $0.775 \pm 0.004$ & $0.783 \pm 0.003$\\
$f_2$ & $0.719 \pm 0.007$ & $0.731 \pm 0.003$\\
$F_1$ & $0.026 \pm 0.002$ & $0.028 \pm 0.004$\\
$F_2$ & $0.757 \pm 0.002$ & $0.747 \pm 0.004$\\
\hline
\end{tabular}
\end{center}
\label{day180_fits_param}
\end{table}

\subsection{Simulating day 196}
\label{S:sim196}

\begin{figure}
\resizebox{\hsize}{!}{\includegraphics[angle=-90]{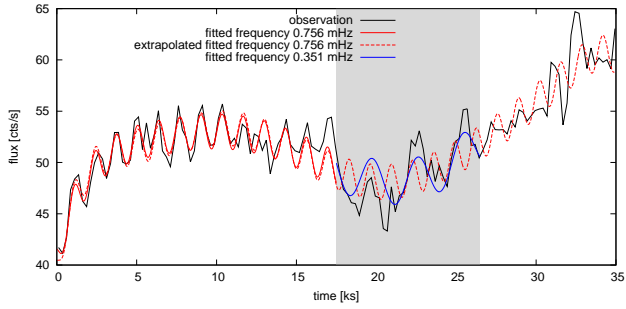}}
\caption{Different sine fits to the observed data from day 196. The data are averaged (every 250\,s) for clarity. The different white-gray areas represent different time intervals used for sine model fitting (see text and Table~\ref{day196_fits_param} for details).}
\label{day196_fits}
\end{figure}

In order to study different possible interpretations of the observed periodogram from day 196, we simulated light curves with the same long term trend\footnote{Obtained by fitting a 7th-order polynomial.} as the observed data but with different modulations. We sampled and noised (using Gaussian noise) a sine function to get a similar scatter as in the observed data. The first step is based on visual inspection of the observed data. We divided the light curve into three different intervals during which the variability differs (Fig.~\ref{day196_fits}). The first interval is from 0 to 17.5\,ks where the periodic variability is clearest. The next interval ranges from 17.5 to 26.5\,ks and still shows some periodic variability, but with a possibly lower frequency (although we only have 2-3 cycles). After 26.5\,ks, the periodic oscillations disappear, only some patterns are visible toward the end of the light curve but hardly recognisable whether it is a coherent or stochastic oscillation. This behaviour can also be recognised in Fig.~\ref{time_map_day196}.

The second step is based on a sine function fitted to the data using the {\tt Gnuplot}\footnote{http://www.gnuplot.info/} software. We fitted the first interval (0 - 17.5 ks) using as start parameters the frequencies of 0.730, 0.748 and 0.764\,mHz that were derived by \citet{dobrotka2010}. All three cases converged to a value of 0.756\,mHz. The second interval (17.5 - 26.5\,ks) yields a lower frequency of 0.351\,mHz. All fitted parameters are summarised in Table~\ref{day196_fits_param}.
\begin{table}
\caption{Fitted sine parameters in different light curve segments from day 196, where $a$ is the amplitude, $f$ is the frequency and $\phi$ is the phase in the sine argument $2 \pi t f + \phi$ ($t$ is time). The errors are standard errors calculated from the covariance matrix by the {\tt GNUPLOT} software.}
\begin{center}
\begin{tabular}{cccc}
\hline
\hline
fitted int. & $a$ & $f$ & $\phi$\\
(ks) & (counts/s) & (mHz) &\\
\hline
0 - 11.0 & $1.694 \pm 0.250$ & $0.756 \pm 0.005$ & $2.618 \pm 0.287$\\
17.5 - 26.5 & $2.136 \pm 0.455$ & $0.351 \pm 0.012$ & $2.010 \pm 1.698$\\
\hline
\end{tabular}
\end{center}
\label{day196_fits_param}
\end{table}

We performed 100000 simulations using the two sine modulations on the corresponding time intervals from Table~\ref{day196_fits_param} and computed a goodness parameter for each simulation that assesses the agreement of the power spectrum derived from the simulated light curve with the observed power spectrum. We choose the sum of unweighted squared residua, $\Sigma = \sum_{i=0}^{n}(m(f_i)-d(f_i))^2$, where $m(f_i)$ and $d(f_i)$ are model and data at frequency $f_i$, respectively. We computed $\Sigma$ probing two frequency intervals of $f_i=0.713..0.784$\,mHz and $f_i=0.680..0.819$\,mHz. The former takes into account only the two dominant peaks, while the latter also includes the two small satellite peaks at approximately 0.7 and 0.8\,mHz.

\begin{figure*}
\resizebox{\hsize}{!}{\includegraphics[angle=-90]{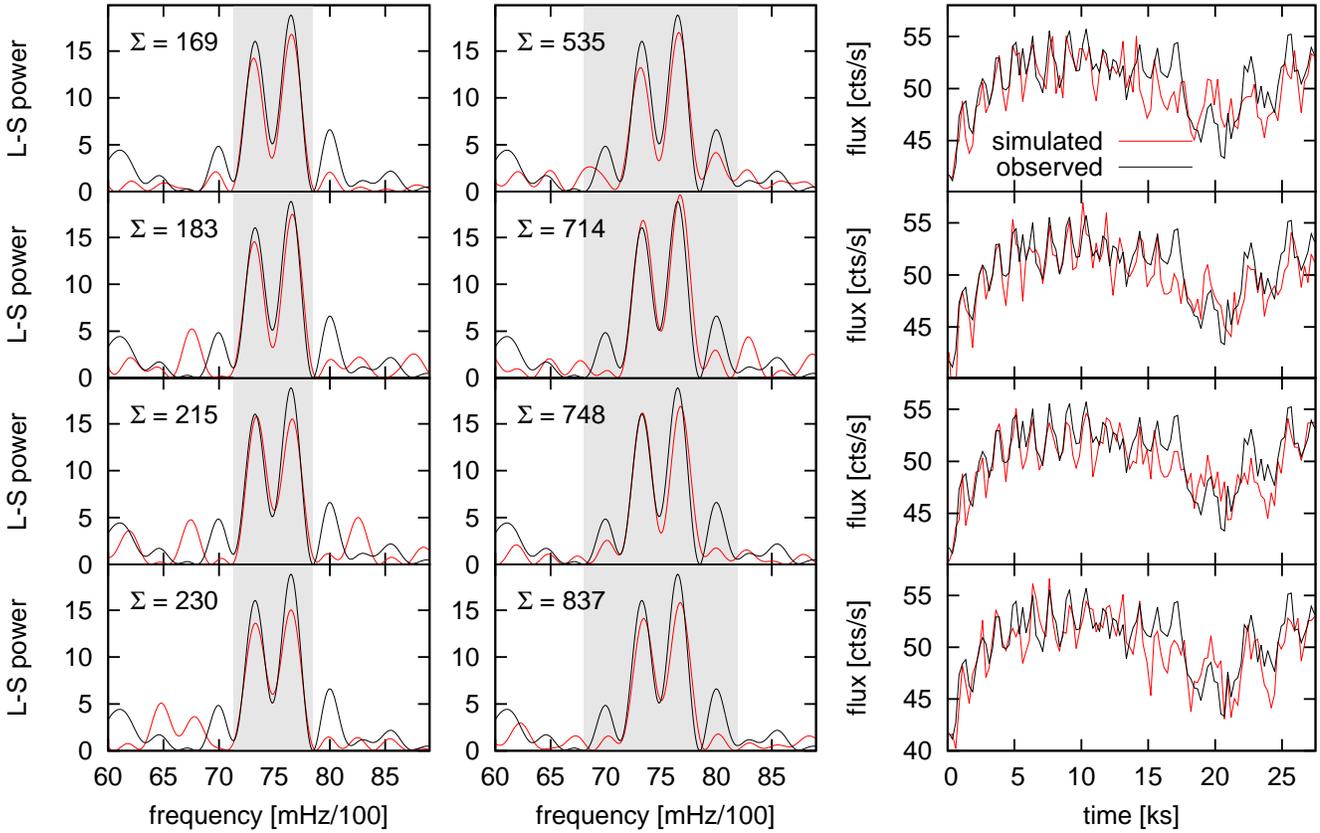}}
\caption{The 4 best cases selected out of 100000 simulations described in Sect.~\ref{S:sim196}, compared to the data taken on day 196. Left and middle column: simulated (red) and observed (black) periodograms with $\Sigma$ in respective upper left corners, using two different frequency intervals that are shown as gray-shaded areas. Right column: corresponding simulated (red) light curves compared to observed light curve (black) belonging to the PSD shown in the respective middle panels.}
\label{best_simul}
\end{figure*}

\begin{figure}
\resizebox{\hsize}{!}{\includegraphics[angle=0]{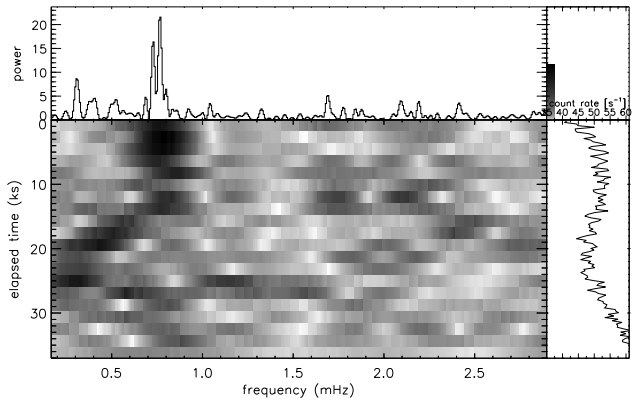}}
\caption{Time map of the best simulated light curve depicted in the top right panel of Fig.~\ref{best_simul}.}
\label{time_map_best_simul}
\end{figure}

Out of all simulations, we select the 4 best ones for illustration in Fig.~\ref{best_simul}. The left column shows the best fits (from top to bottom) based on the narrow frequency interval while the middle column shows the best simulations based on the broader frequency interval. The light curves used for the middle panel calculation are shown in the right column. Clearly a single frequency is sufficient to generate the observed main feature at 0.75\,mHz and no multiple signals as suggested by \citet{leibowitz2006} and \citet{dobrotka2010} are needed. Moreover, the signal is present or dominant in the interval 0 - 17.5\,ks. The rest of the light curve is dominated by longer time scale modulations or just noise without any significant periodic signal. The time map corresponding to the best simulated light curve is depicted in Fig.~\ref{time_map_best_simul}. Similar patterns can be seen. Worth noting is the fake smooth frequency decrease below 20\,ks from 0.756 to 0.351\,mHz resulting from the abrupt change in modulating frequencies.

We also performed simulations assuming a single signal in the full light curve and with a constant amplitude. In such a case only a single and dominant peak is detected. This supports the connection between the double peak feature and variable modulation amplitude, i.e. that the signal is not present in the whole light curve.

Moreover, we also performed some test simulations assuming a variable frequency over different time intervals. The resulting periodograms showed a multiple peak structure also in the case in which the variability was present in the full light curve with constant amplitude. However, the pattern was very different, unacceptable and shifted toward lower frequencies or showed strong aliases not present in the observed periodogram. Moreover, the resulting sum of $\Sigma$ was considerably worse than in the constant frequency case. To get periodograms in better agreement with the observations we needed to restrict the modulation only to the interval of 0-17.5\,ks, i.e. a variable amplitude is needed. Therefore, the double peak was probably again the result of a false beating rather than a variable frequency.

\subsection{Variability profile}

\begin{figure}
\resizebox{\hsize}{!}{\includegraphics[angle=-90]{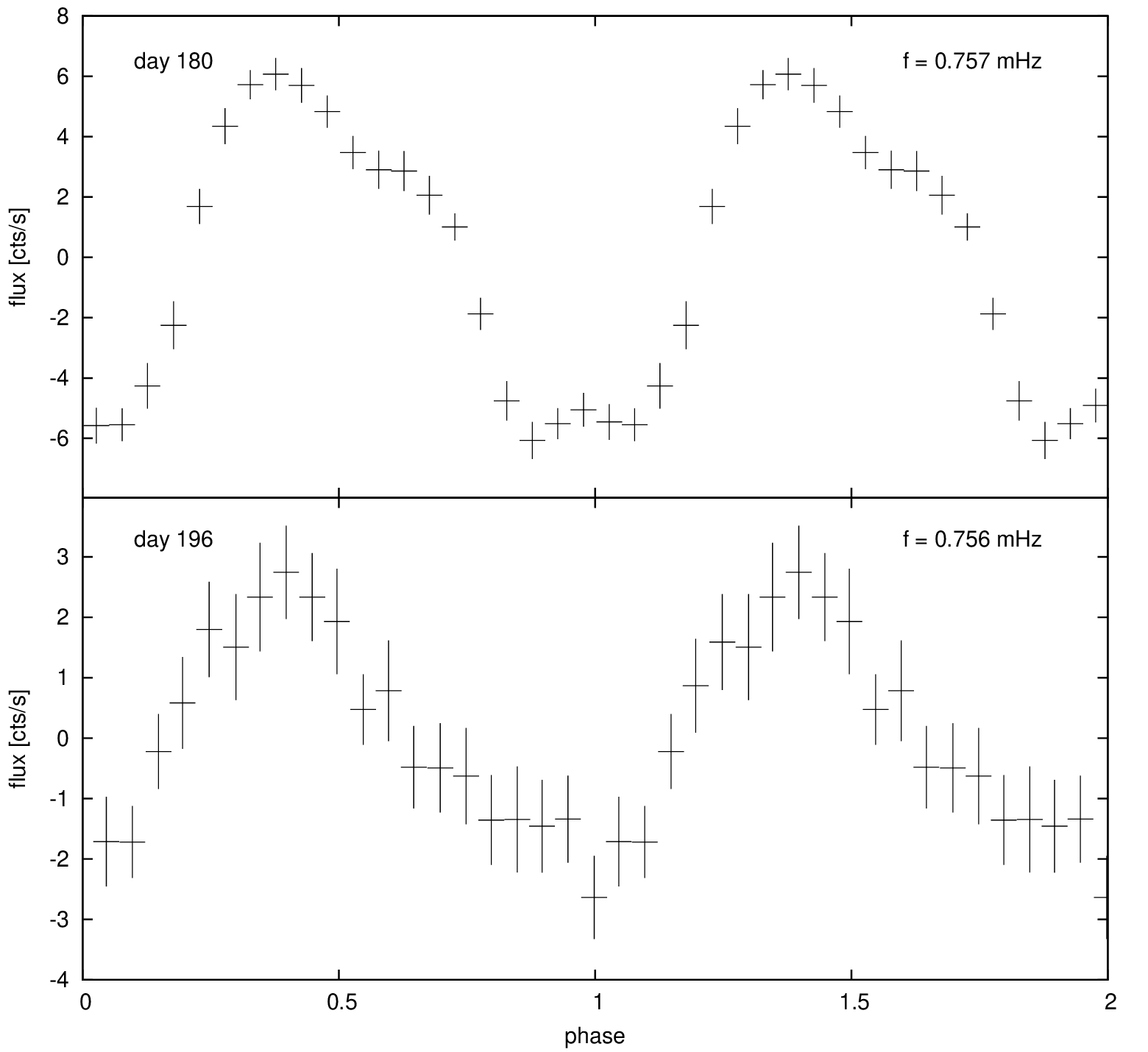}}
\caption{Folded and binned light curves subsample between 0 and 17\,ks on day 180, and 0 and 11\,ks on day 196. The folding frequency is marked in each panel. The error-bars are errors of the mean and two cycles are shown for clarity.}
\label{folded_lc}
\end{figure}

To study the profile of the modulation, we used detrended light curves. We folded the first 17\,ks of the day 180 light curve using the frequency of 0.757\,mHz from Table~\ref{day180_fits_param}, and the first 11\,ks of the day 196 light curve, during which the variability is most obvious, using the fitted frequency of 0.756\,mHz from Table ~\ref{day196_fits_param}. The folded and binned light curves are depicted in Fig.~\ref{folded_lc}. Apparently, the profile on day 180 shows an obviously non sinusoidal profile with a constant plateau at the minimum. The profile on day 196 shows a more or less triangular and asymmetric shape of the pulse. Both profiles are similar, except that day 196 does not show the obvious hump at phase 0.7 as in day 180.

Following Fourier series theory, a non-sinusoidal shape must yield harmonic signals in the periodogram. This is confirmed by the clear presence of overtones in Figs.~\ref{time_map_day180} and \ref{time_map_day196} already mentioned in \citet{ness2003}. Since our simulations are based on pure sine functions, the overtones are not present in the simulated time maps (Fig.~\ref{time_map_best_simul}). The exact variability pattern depends on the shape of the accretion spot on the white dwarf surface, and a non-sinusoidal variability profile is common in intermediate polars (see e.g. \citealt{beardmore1998}).

\section{Discussion}

The previous works about nova V4743\,Sgr suggested that two X-ray observations taken 180 and 196 after the maximum brightness were modulated with two signals with close frequencies (\citealt{leibowitz2006}, \citealt{dobrotka2010}). We now show that this result does not account for the possibility of variations during these observations which can manifest themselves as multiple signal in a time-averaged periodogram. The review of the mentioned observations in this paper and use of basic mathematics together with simulations reveals that a single signal can explain at least equally well the observed behaviour.

\subsection{Day 180}
\label{discussion_day180}

\citet{dobrotka2010} showed with a two-component fitting of the \chandra\ observation from day 180, that the data can be well described by two close frequencies despite the fact, that the standard periodogram showed only one single but broad peak. In this paper we have shown, that this solution is only one possible interpretation, while an additional and totally equivalent solution exists because the ambivalent nature of the underlying mathematical description of two summed sine functions. According to this second solution, the data are modulated by a single and constant frequency, while the modulation amplitude varies with another much shorter frequency. This result is a sort of beating, which is also a natural consequence when two sine functions with close frequencies are summed. Therefore, both solutions are inherently equivalent and indistinguishable. However, the nova experienced a strong flux decline during the observation on day 180 and the variability amplitude decreased\footnote{However, following Fig.~2 in \citet{dobrotka2010} the fractional amplitude is not affected.} simultaneously with the flux. Such amplitude decrease can mimic the mentioned beating. Therefore, the solution where the amplitude decreases can be generated by the flux decline which appears more probable to us. Another solution (and harder to believe) could be that the amplitude decrease is due to real beating of two signals that conspicuously coincide with the flux decline.

Moreover, the derived constant frequency has a value of $0.757 \pm 0.002$\,mHz (Table~\ref{day180_fits_param}) which agrees within the errors with the white dwarf rotation frequency detected as stable signal 2 and 3.5 years after maximum in \citet{dobrotka2010}. Note that the X-ray observations taken 2 and 3.5 years after maximum are much longer, leading to a well-constrained frequency.

\subsection{Day 196}

\subsubsection{Main signal}

Furthermore, we showed that the double periodogram peak of the main signal around 0.76\,mHz detected in the \xmm\ observation taken 196 days after outburst can be an artefact with a single frequency being sufficient to explain this observation (double peak). The signal is not present during the entire observation and a simple visual inspection suggests that the variability is only present during the first 17.5\,ks. After 17.5\,ks the light curve passes a minimum with a constant and strong increase of brightness toward the end with no obvious uniform modulations, only some lower frequency (0.351\,mHz) peaks are observed between 17.5 and 26.5\,ks.

\begin{figure}
\resizebox{\hsize}{!}{\includegraphics[angle=-90]{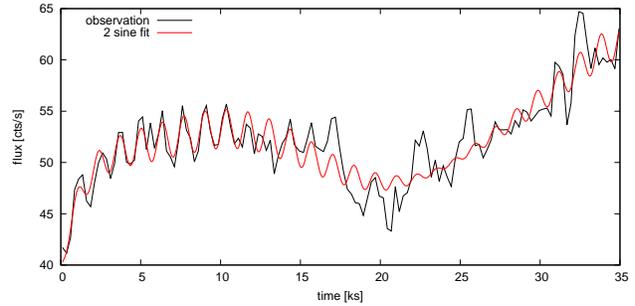}}
\caption{Fit of the day 196 observation using superposition of two sine functions with close frequencies and the visualisation of the resulting beating (see text for details). The light curve data are averaged (every 250\,s) for clarity.}
\label{run3_2sine_fit}
\end{figure}

\citet{dobrotka2010} used a two-dimensional periodogram\footnote{The data were fitted with two sine functions with frequencies as input mesh parameters in order to calculate the residuals, i.e. a 2 dimensional equivalent of a standard periodogram.} in order to study the multifrequency nature of the main 0.75\,mHz feature. Their finding is in agreement with the Lomb-Scargle periodogram, i.e. the data are modulated with two signals with close frequency values. Therefore, the observation must have the characteristics described in Section~\ref{fitting_day_180} and based on equation (\ref{sine_equality_1}), i.e. a beating should appear in the data. However, the disappearance of the main 0.75\,mHz signal after approximately 17.5\,ks is identified by the fitting algorithm as amplitude minimum which mimics the knot of a beat cycle (Fig.~\ref{run3_2sine_fit}). The appearance of two peaks between 30 and 35\,ks sustain this potentially false "interpretation" mimicking a second beating cycle. Therefore, an optimisation algorithm converges to a beating solution, falsely resulting in two superposed signals with two close frequencies, while the amplitude decrease/disappearance of the main variability around 17.5\,ks can have a totally different origin just as on day 180. This explains why the power spectrum from the full light curve yields two close peaks.

Finally, just as for the day 180 we conclude that the fitted frequency of $0.756 \pm 0.005$\,mHz agrees well with the white dwarf rotation frequency derived previously by \citet{dobrotka2010}.

\subsubsection{Other signals}

As already suggested by \citet{leibowitz2006} and confirmed by \citet{dobrotka2010} other signals are present in the light curve, supporting the presence of other forms of pulsation. In this work we mentioned an additional and obvious frequency of $0.35 \pm 0.01$\,mHz (blue line in Fig.~\ref{day196_fits}) which is close to the value of 0.38\,mHz listed in Table~1 in \citet{leibowitz2006}.

It is worth noting that the shorter frequency of 0.35\,mHz appeared at the brightness decrease at $\sim 17$\,ks into the observation at the same time as main frequency of 0.76\,mHz disappeared. This can be a result of some physical processes during which the observable emission experienced a temporary reduction, accompanied by some types of pulsation, while the radiation from the accretion flow on the white dwarf surface vanished or weakened considerably making the main signal disappear. Possible processes that could lead to such a reduction in brightness could be obscurations or photospheric expansion.

Visual inspection of the Fig.~\ref{day196_fits} suggests that the low frequency peaks (blue line) are coherent in phase with alternate peaks of the high frequency modulation (red line). This would suggest that the lower frequency of 0.35\,mHz above 17\,ks is the same modulation bellow 17\,ks but continuing with alternate peaks vanishing. The red dashed line is the extrapolation of the modulation bellow 17\,ks and it is clear that the low frequency peaks are not in phase. Therefore, this supports the interpretation of an independent variability source.

\subsection{Implications}

The main feature around 0.75\,mHz has two possible interpretations, either pulsations or modulations excited by the rotation of the white dwarf (\citealt{dobrotka2010}). The double-peak solution of the main signal would be a strong counter argument against white dwarf rotation, but the new interpretation of a single signal that is variable in amplitude revitalises the interpretation of rotation. The same period was also seen, as a single signal, on days 302, 371 and 526, thus in all observations taken during the SSS phase. While pulsations might occur during the SSS phase one would expect them to disappear when the nova turns off, but the same period was also detected in the quiescent light curves on days 742 and 1286. If pulsations of 0.75\,mHz occur in the quiescent white dwarf, one would expect it to be disturbed or damped within the expanding nova ejecta during the outburst, but detecting the same pulsation period both during and after the nova outburst would be a challenge. We thus strongly prefer the interpretation of a magnetic white dwarf in an intermediate polar whose rotation stimulates observable periods.

This paper has also some technical implications. Detailed period analysis of rather short satellite data should be taken with caution, and any standard periodogram study should be accompanied with simulations or more detailed treatment in order to investigate any false beating. The same amplitude variation, probably generated by a flux dip, is also noticeable in the X-ray light curve of V2491\,Cyg, yielding a double-peak feature as a main signal in the corresponding periodogram (\citealt{ness2011}). The presence of the first harmonics in the first part of the observation played also a significant role. The false beating was proposed as a probable reason of the double peak nature. This was confirmed by simulating light curves with a single signal and the amplitude as observed, while a double peak emerged in the corresponding periodograms. Apparently dips, flux declines, variable amplitudes are common in novae during outburst. Therefore, any multiple peak detection in this stage of nova evolution is an indicator for an instable signal that deserves a closer look with time mapping techniques as shown in Fig.~\ref{time_map_day180}.

\section{Summary and conclusions}
\label{summary}

The results of this work can be summarised as follows:

(i) The X-ray observation taken 180 days after maximum brightness has two inherently indistinguishable interpretations: 1) the data are modulated with two signals with close frequency values, 2) a single and constant frequency is present, but the periodogram is influenced by the strong flux decrease and a contemporaneous reduction of the modulation amplitude.

(ii) Simulations reveal that also the double peak nature of the main 0.75\,mHz feature during the X-ray observation taken on day 196 can be explained by a single signal, if it is only present during the first 17\,ks. The double nature of the main peak in the full power spectrum is then the result of a fake beating because the period is only present during a fraction of the whole light curve.

(iii) Other signals are present as summarised by \citet{leibowitz2006}, but some of them can have a similar explanation as the double nature of the main signal, i.e. fake multiple nature due to fake beating.

(iv) The false double peak nature of the signal during a nova outburst is not unique for V4743\,Sgr. It has also been detected in X-ray data of the nova V2491\,Cyg during outburst. A false beating generated by variability amplitude change was also proposed as a reason in that case.

The main conclusion of this work is that the new finding about the main signal together with the folded variability profile converges toward an intermediate polar interpretation rather than pulsations.

\section*{Acknowledgements}

AD was supported by the Slovak grant VEGA 1/0335/16 and by the ERDF - Research and Development Operational Programme under the project "University Scientific Park Campus MTF STU - CAMBO" ITMS: 26220220179.

\bibliographystyle{mn2e}
\bibliography{mybib}

\end{document}